\pdfoutput=1

\documentclass[11pt]{article}

\usepackage[final]{acl}

\usepackage{times}
\usepackage{latexsym}
\usepackage[T1]{fontenc}
\usepackage[utf8]{inputenc}
\usepackage{microtype}
\usepackage{inconsolata}

\usepackage{array}
\usepackage{makecell}

\usepackage[TABBOTCAP]{subfigure}
\usepackage[shortlabels]{enumitem}
\usepackage{placeins}
\usepackage{bbold}
\usepackage{tikz-dependency}
\usepackage{algorithm}
\usepackage{algpseudocode}
\usepackage{multirow}
\usepackage{booktabs}

\usepackage{color}
\usepackage{helvet}
\usepackage{booktabs}

\usepackage{xcolor}
\usepackage{bbm}
\usepackage{graphicx}
\graphicspath{ {images/} }
\usepackage{amsmath}
\usepackage{amsfonts}

\usepackage{hyperref}
\usepackage{url}
\usepackage{tabularx}
\usepackage{xspace}
\usepackage{comment}

\usepackage{algorithm}
\usepackage{algpseudocode}
\algrenewcommand\algorithmicindent{1em}
\usepackage{listings}
\usepackage{booktabs}

\newcommand{\ourwork}{\textsc{cAST}\xspace}

\usepackage{color, colortbl}
\definecolor{amber}{rgb}{1.0, 0.75, 0.0}
\definecolor{applegreen}{rgb}{0.55, 0.71, 0.0}
\definecolor{treegreen}{rgb}{0.13, 0.54, 0.23}
\definecolor{LightCyan}{rgb}{0.88,1,1}

\newcommand{\good}{\cellcolor{applegreen!15}}
\newcommand{\better}{\cellcolor{applegreen!40}}

\usepackage{xcolor}
\usepackage{color-edits}
\addauthor{sw}{purple}

\title{\ourwork: Enhancing Code Retrieval-Augmented Generation \\ with Structural Chunking via Abstract Syntax Tree
}

\author{
 \textbf{Yilin Zhang$^1$}\thanks{\quad Corresponding contact email addresses: \{jasonzh3,sherryw\}@andrew.cmu.edu. Our code is available at \url{https://github.com/yilinjz/astchunk}}\quad
 \textbf{Xinran Zhao$^1$}\quad
 \textbf{Zora Zhiruo Wang$^1$}\quad
 \textbf{Chenyang Yang$^1$}\quad \\
 \textbf{Jiayi Wei$^2$}\quad
  \textbf{Tongshuang Wu$^1$}\\
 $^1$Carnegie Mellon University, $^2$Augment Code
}
\date{}

\begin{document}
\maketitle
\begin{abstract}
\label{sec:abstract}
Retrieval-Augmented Generation (RAG) has become essential for large-scale code generation, grounding predictions in external code corpora to improve factuality. 
However, a critical yet underexplored aspect of RAG pipelines is chunking---the process of dividing documents into retrievable units.
Existing line-based chunking heuristics often break semantic structures, splitting functions or merging unrelated code, which can degrade generation quality. 
We propose chunking via Abstract Syntax Trees (\ourwork), a structure-aware method that recursively breaks large AST nodes into smaller chunks and merges sibling nodes while respecting size limits. This approach generates self-contained, semantically coherent units across programming languages and tasks, improving performance on diverse code generation tasks, e.g., boosting Recall@5 by 4.3 points on RepoEval retrieval and Pass@1 by 2.67 points on SWE-bench generation. Our work highlights the importance of structure-aware chunking for scaling retrieval-enhanced code intelligence.

\end{abstract}
\section{Introduction}
\label{sec:introduction}

 \begin{figure}[t!]
    \centering
    \includegraphics[width=\linewidth, trim={0 21cm 43cm 0cm}, clip]{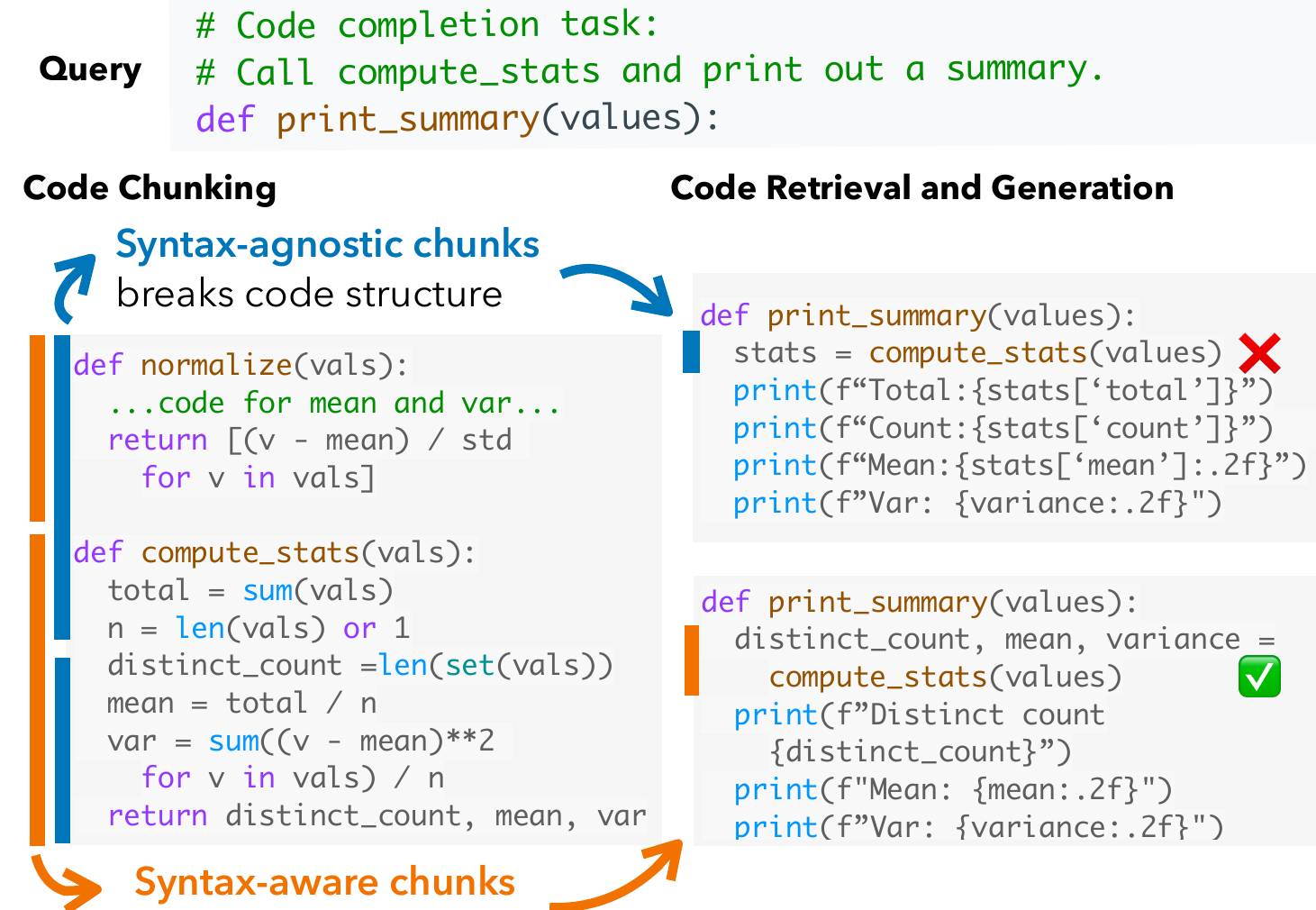}
    \vspace{-10pt}
    \caption{Syntax-agnostic chunking often omits crucial information needed to generate functional code. In this example, fixed-size chunking breaks the structure of the $\texttt{compute\_stats}$ method, causing the model to lose context regarding its return value. As a result, the model generates incorrect code based on a mistaken assumption of what is returned. In contrast, when given syntax-aware chunks, the model accurately identifies the return values and integrates them correctly within the existing codebase.}
    \vspace{-10pt}
\label{fig:teaser}
\end{figure}

Large‑scale code generation has emerged as a cornerstone of modern software engineering, powering tasks that range from automated bug fixing \citep{meng2024-llm_bug_fixing} to full‑fledged repository-level completion \citep{zhang2023-repo_code_completion}. 
Retrieval‑augmented generation (RAG) pushes this frontier further by allowing language models to ground their predictions in a rich external corpus of data \citep{guu2020retrieval}, effectively mitigating hallucinations and improving factual correctness \citep{Izacard2022FewshotLW}.

One crucial preprocessing step in Retrieval-Augmented Generation (RAG) is chunking~\cite{bohnet2023attributedquestionansweringevaluation}—breaking large documents into manageable segments that can be efficiently indexed, retrieved, and used as contextual input during generation.
To date, most chunking approaches rely on fixed-size, line-based splitting~\cite{lewis2020rag}. While simple and generally effective, this method struggles with structured content like code, where the document naturally contains semantic or syntactic blocks. As shown in Figure~\ref{fig:teaser}, naive chunking often splits meaningful units (e.g., functions and classes) across different chunks, losing structural integrity and context.

Can we chunk documents more intelligently, preserving their original structure?
In this work, we explore \ourwork---Chunking via Abstract Syntax Trees. ASTs represent code as hierarchical trees with typed nodes corresponding to program units. By parsing source code into an AST, we apply a recursive, split-then-merge algorithm to convert tree structures into chunks that are better aligned with syntactic boundaries.

Extensive experiments show that \ourwork improves performance across a range of code generation tasks. Specifically, it offers three key advantages:
(1) \emph{Structure-preserving chunks}: AST traversal yields more self-contained chunks, improving both retrieval and generation. For instance, StarCoder2-7B sees an average of 5.5 points gain on RepoEval~\citep{zhang-2023-repoeval}.
(2) \emph{Cross-language consistency}: The language-agnostic nature of \ourwork 
enables better generalization across programming languages, achieving up to 4.3 points gain on CrossCodeEval \citep{ding-2023-crosscodeeval}. 
(3) \emph{Metadata retention}: AST-based chunks more faithfully capture metadata at the file, class, and function levels, enhancing context matching in hybrid code+natural language tasks, e.g., up to 2.7 points gain on SWE-bench \citep{jimenez-2024-swebench}, which focuses on resolving GitHub issues.

\section{\ourwork}

We focus on the first stage of the RAG pipeline: \emph{chunking}. In this step, source code is parsed into semantically meaningful units (such as functions or classes) while preserving the structure of the code. 
These units are then grouped into coherent chunks, which serve as the retrievable context that can be obtained by a subsequent \emph{retriever} and used to prompt a \emph{language model}.

\paragraph{Design Goal.}

Our design for \ourwork\ pursues four aligned goals: (1) \emph{syntactic integrity}---whenever possible, chunk boundaries should align with complete syntactic units instead of splitting them; (2) \emph{high information density}---each chunk is packed up to, but not beyond, a fixed size budget to maximize content utility; (3) \emph{language invariance}---the algorithm employs no language-specific heuristics so it works unchanged across diverse programming languages and code-related tasks; and (4) \emph{plug-and-play compatibility}---concatenating the chunks must reproduce the original file verbatim, enabling seamless drop-in replacement within existing RAG pipelines.

\paragraph{AST Parsing.}
To support syntax-aware chunking, we leverage the \emph{Abstract Syntax Tree (AST)} representation of code.
An AST is a tree-structured abstraction that captures the syntactic structure of source code in a way that is both hierarchical and semantically rich. Rather than treating code as plain text, AST encodes language constructs---like functions, classes, loops, and conditionals---as distinct nodes in a structured parse tree. This enables us to identify meaningful code boundaries with precision, ensuring that chunking respects the underlying syntax. 
Since ASTs are widely supported across languages, this approach also enhances the language-invariance and portability of our method.
Our work uses the \texttt{tree-sitter} library~\citep{treesitter} for the AST tree parsing.

 \begin{figure*}[t!]
    \centering
    \includegraphics[clip,trim={0cm 0cm 0cm 0cm},width=0.98\linewidth]{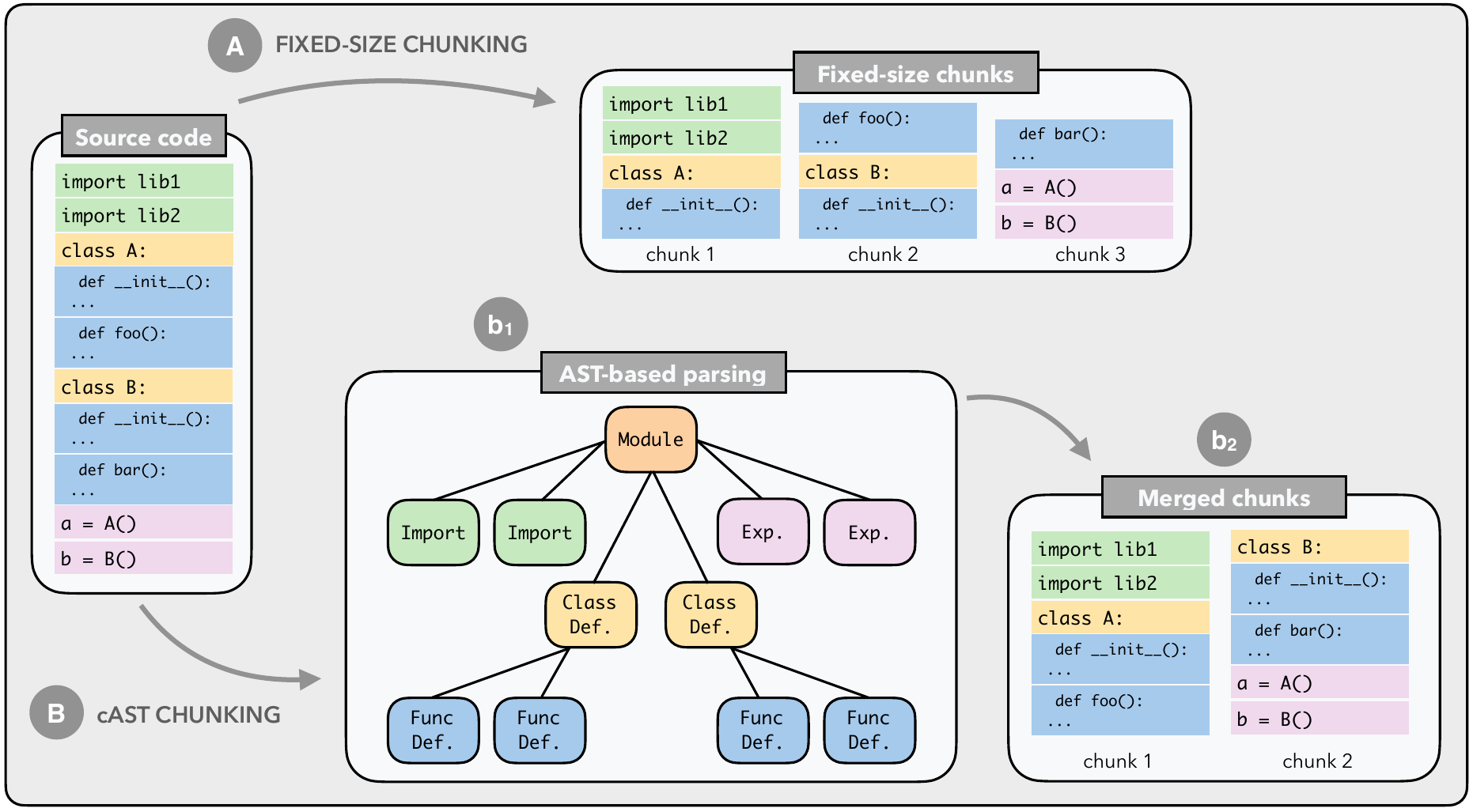}
    \caption{Comparison of fixed-size chunking vs. \ourwork. For \ourwork, we first parse the document into a tree of AST nodes. Then, starting from the first level, we greedily merge AST nodes into chunks. If adding a node would exceed the chunk size limit, we recursively break it into smaller nodes. The output of \ourwork is a list of chunks where each chunk contains a list of AST nodes.}
    \label{fig:ast_structure}
    \vspace{-0.1in}
\end{figure*}

\paragraph{AST-based Recursive Chunking.} 
With the AST tree at hand, we use a recursive, split-then-merge algorithm for converting tree structures into chunks, as shown in Figure~\ref{fig:ast_structure}.
To retain as much syntactic information as possible, we first traverse the tree in a top-down manner, to fit those large AST nodes into a single chunk whenever possible. For those nodes that must be split due to exceeding the chunk size limit, to avoid too many overly small chunks, we further perform a greedy merging step, combining adjacent small sibling nodes into one chunk, to maximize the per-chunk information density. The detailed process is also described in Alg.~\ref{alg:cast}.

\paragraph{Chunk size metric.}
Choosing an appropriate budget for each chunk is nontrivial: two segments of equal line count can carry wildly different amounts of code, and AST-aligned chunks naturally vary in their physical span (e.g., a single import line versus an entire class body). So unlike prior work~\cite{wang2024coderagbench}, we measure chunk size by the number of non-whitespace characters rather than by lines. This keeps chunks text-dense and comparable across diverse files, languages, and coding styles, ensuring that our budget reflects actual content rather than incidental formatting. 

\section{Experiments}
\label{sec:experiment}

We evaluate \ourwork with various top retrieval and generation models in various code task settings. We present results of selected end-to-end RACG pipelines (retriever + LM) in Section \ref{subsec:RACG_Performance} and full tables in the Appendix (\ref{tab:full_retrieval_results}, \ref{tab:full_repoeval_results}, \ref{tab:full_swebench_results}, \ref{tab:full_cceval_results}).

\subsection{Experiment Settings}
\label{subsec:exp_setup}

\paragraph{Datasets.} We evaluate \ourwork on various software engineering (SE) tasks using three benchmarks: 
\begin{itemize}[labelwidth=*,leftmargin=1.3em,align=left, topsep=0pt, nosep]
    \item RepoEval~\citep{zhang-2023-repoeval}: Code completion tasks with long intra-file contexts;
    \item CrossCodeEval~\cite{ding-2023-crosscodeeval}: Multi-language queries requiring cross-file reasoning;
    \item SWE-bench~\cite{jimenez-2024-swebench}: General SE tasks involving code patch generation. We use the SWE-bench Lite variant~\citep{swebenchlite}, a 300-problem subset where each issue is solvable by editing a single file.
\end{itemize}

\paragraph{Metrics.} For retrieval performance, we report three common metrics:
nDCG, Precision and Recall, with $k=5$. Notably, since retrieval scores from different corpus distributions are not directly comparable, we implement a score mapping technique to align AST-based retrieval scores with those of the baseline, with details in Appendix~\ref{sec:line_mapping_metric}.

As for generation, we use Pass@k \citep{chen2021evaluatinglargelanguagemodels} for execution-based datasets and match-based metrics for the others, following prior work \citep{wang2024coderagbench, ding-2023-crosscodeeval}. 
Specifically, we report the canonical Pass@1 score for RepoEval and SWE-bench. Additionally, we record the Pass@8 score for SWE-bench by sampling multiple responses with high temperature following Agentless \citep{xia2024-agentless} to examine the robustness of \ourwork. 
For CrossCodeEval, we report exact match (EM), edit similarity (ES), and other identifier match metrics in the original work.

\paragraph{Retrieval and Generation Models.}
We adopt various kinds of retrievers, including general-text dense retrievers: \texttt{BGE-base}~\citep{bge_embedding} and \texttt{GIST-base}~\citep{solatorio2024gistembed}; and code-specific retriever: \texttt{Codesage-small-v2}~\citep{zhang2024codesage}, following CodeRAG-Bench \citep{wang2024coderagbench}.


Similarly, for generations, we include two code-specific LMs: \texttt{StarCoder2-7B}~\citep{lozhkov2024starcoder}, \texttt{CodeLlama-7B-Python}~\citep{roziere2023code}; and two general-purpose LMs (\texttt{claude-3.7-sonnet}, \texttt{gemini-2.5-pro-0325}), as both represent the state-of-the-art in coding.

Further details of our experimental setup are introduced in Appendix~\ref{sec:experiment_setup}.

\subsection{\ourwork Results and Analysis}
\label{subsec:RACG_Performance}
Table~\ref{tab:cognac_main_table} presents the end-to-end RACG results with selected retrievers (BGE-base, GIST-base, Codesgae-small-v2) on the three datasets. The results highlight several key observations:

\newcommand{\metricmodel}[2]{#1 {\small (\texttt{#2})}}

\begin{table}[t!]
\small
\centering
\setlength{\tabcolsep}{2pt}
\resizebox{\linewidth}{!}{
  \begin{tabular}{l |l|ccc|ccc}
    \toprule
    & \multicolumn{1}{c|}{\multirow{2}{*}{\textbf{Metric (Model)}}} & \multicolumn{3}{c|}{\textbf{\ourwork chunking}} & \multicolumn{3}{c}{\textbf{Fixed-size chunking}} \\
    \cmidrule(lr){3-5} \cmidrule(lr){6-8}
    {} & & {BGE} & {GIST} & {CodeSage} & {BGE} & {GIST} & {CodeSage}\\
\midrule

\multicolumn{8}{c}{RepoEval} \\
\midrule
\multirow{3}{*}{R}
    & nDCG & 71.1 & \good{75.9} & \good{85.1} & 71.3 & 74.2 & 83.0\\
    & Precision  & \good{34.9} & \better{38.1} & \good{44.1} & 32.8 & 34.8 & 42.9\\
    & Recall  & \good{69.8} & \better{75.0} & \good{83.9} & 67.4 & 70.7 & 82.1\\
\midrule
\multirow{2}{*}{G}
    & \metricmodel{Pass@1}{StarCoder2} & \better{51.7}  & \better{57.9} & \better{73.2} & 47.5 & 51.2 & 67.6\\
    &  \metricmodel{Pass@1}{CodeLlama} & \better{49.6} & \better{56.6} & \better{72.1} & 45.6 & 51.5 & 66.5\\
\midrule\midrule

\multicolumn{8}{c}{SWE-Bench} \\
\midrule
\multirow{3}{*}{R}
    & nDCG & \good{44.0} & \good{44.4} & 43.1 & 42.4 & 43.1 & 42.6\\
    & Precision  & \good{39.7} & 39.1 & \good{38.8} & 38.3 & 38.6 & 37.5\\
    & Recall  & \good{18.4} & 18.5 & 18.3 & 17.3 & 17.8 & 17.5\\

\midrule
\multirow{2}{*}{G}
    & \metricmodel{Pass@1}{Claude} & \better{16.3} & 15.0 & \better{16.7} & 13.7 & 14.7 & 14.0 \\ 
    &  \metricmodel{Pass@8}{Gemini} & \better{35.3} & 33.7 & \better{32.7} & 32.3 & 33.0 & 31 0 \\
\midrule\midrule


\multicolumn{8}{c}{CrossCodeEval} \\
\midrule
\multirow{1}{*}{R}
    & Identifier Match (EM) & \good{34.7} & 34.0 & \better{39.9} & 32.0 & 33.5 & 36.3\\

\midrule
\multirow{2}{*}{G}
    & \metricmodel{EM}{StarCoder2} & \good{23.8} & 23.4 & \better{29.1} & 21.2 & 23.0 & 24.8\\
    &  \metricmodel{ES}{StarCoder2} & \good{72.2} & 71.9 & \good{74.3} & 71.0 & 71.7 & 73.1\\

\bottomrule
  \end{tabular}
}
\vspace{-1mm}
\caption{\textbf{R}etrieval and \textbf{G}eneration Performances across three benchmarks, using different retrieval models (BGE, GIST, CodeSage) and different LMs (full model names in \S\ref{subsec:exp_setup}).}
\label{tab:cognac_main_table}
\vspace{-0.2in}
\end{table}

\paragraph{Retrieval.}
\ourwork’s structure-aware chunking steadily improves retrieval performance across datasets and retrievers. Specifically, all models show gains of 1.2–3.3 points in Precision and 1.8–4.3 in Recall on code-to-code retrieval (RepoEval), and 0.5–1.4 in Precision and 0.7–1.1 in Recall on the more challenging NL-to-code retrieval (SWE-Bench). These improvements suggest that aligning chunks with abstract syntax boundaries helps diverse retrievers surface semantically coherent code fragments, supplying richer and more accurate evidence for downstream tasks.

\paragraph{Generation.}
\ourwork benefits both intra-file and cross-file code completion. Notably, gains are most pronounced when the RACG pipeline employs code-specific retrievers, implying that the structurally aligned chunks deliver fuller context to both the specialized retriever and the generation model, which in turn facilitates more accurate context retrieval and coherent code synthesis. On NL-to-code generation, we observe remarkable gains with BGE-base and CodeSage retrievers under one and multiple rounds of sampling. 

\paragraph{Correlation between retrieval and generation performance.} Among the three retrieval metrics we use, we notice that higher precision tends to convert into better generation performance, aligning with conclusions from prior work \citep{zhao2024-beyondrelevance}. This suggests that ensuring the top-k context is highly relevant reduces noise and enables the language model to concentrate on concise, accurate evidence, thereby boosting answer fidelity \citep{fang2024-noiserobustness, salemi2024-retrievalquality}.

By contrast, recall-oriented metrics and nDCG correlate only weakly with downstream quality—once the necessary evidence appears in the retrieved set, adding lower-ranked chunks yields diminishing returns or can even hurt performance by introducing distractors.
\section{Ablations}
\label{ablations}
\paragraph{Necessity of merging.} The motivation for introducing merging in our algorithm is to maximize the information density of each chunk. Under a split-only approach, small AST nodes, such as import statements and variable assignments, generate an excessive number of chunks, which unnecessarily enlarges the index and degrades retrieval performance. These fine-grained chunks also contain limited context, making them less effective for downstream tasks, as shown in Table~\ref{tab:cast_ablation_splitting}. Across all retrievers, we find that both retrieval and generation performance decline under the split-only strategy.

\begin{table}[t!]
\small
\centering
\setlength{\tabcolsep}{2pt}
\resizebox{\linewidth}{!}{
  \begin{tabular}{l|l|ccc|ccc}
    \toprule
    & \multicolumn{1}{c|}{\multirow{2}{*}{\textbf{Metric (Model)}}} & \multicolumn{3}{c|}{\textbf{Split-then-merge (\ourwork)}} & \multicolumn{3}{c}{\textbf{Split-only}} \\
    \cmidrule(lr){3-5} \cmidrule(lr){6-8}
    {} & & {BGE} & {GIST} & {CodeSage} & {BGE} & {GIST} & {CodeSage}\\
\midrule
\multirow{1}{*}{R}
    & nDCG & 71.1 & 75.9 & 85.1 & 53.5 & 59.1 & 66.1 \\
\midrule
\multirow{2}{*}{G}
    & \metricmodel{Pass@1}{StarCoder2} & 51.7 & 57.9 & 73.2 & 48.3 & 45.0 & 65.4 \\
    &  \metricmodel{Pass@1}{CodeLlama} & 49.6 & 56.6 & 72.1 & 47.2 & 48.5 & 58.4 \\
\bottomrule
  \end{tabular}
}
\vspace{-1mm}
\caption{Ablation study comparing performance metrics for Split-then-merge (\ourwork) and Split-only methodologies across different models.}
\label{tab:cast_ablation_splitting}
\end{table}


\paragraph{Selection of context length.} In our experiments, we set \textit{max\_context\_length} = 4000, which roughly corresponds to the top five chunks. A comparison of different context lengths is shown in Table~\ref{tab:cast_ablation_context_length}. We observe that doubling the context length does not necessarily improve generation, whereas a modest reduction in context length can lead to performance degradation, likely due to chunk truncation.

\begin{table}[t!]
\small
\centering
\begin{tabular*}{\linewidth}{@{\extracolsep{\fill}}l|l|ccc}
    \toprule
    & \multicolumn{1}{c|}{\multirow{2}{*}{\textbf{Pipeline (R + G)}}} & \multicolumn{3}{c}{\textbf{Context length (tokens)}} \\
    \cmidrule(lr){3-5}
    {} & & {3500} & {4000} & {8000} \\
\midrule
    \multirow{3}{*}{} 
    & BGE + StarCoder2      & 46.9 & 51.7 & 51.7 \\
    & GIST + StarCoder2     & 57.1 & 57.9 & 58.2 \\
    & CodeSage + StarCoder2 & 70.5 & 73.2 & 69.2 \\
\bottomrule
\end{tabular*}
\caption{Ablation study evaluating the impact of different context lengths on the overall performance of several retrieval and generation pipelines.}
\label{tab:cast_ablation_context_length}
\end{table}



\paragraph{Selection of maximum chunk size.} We set \textit{max\_chunk\_size} = 2000 in our experiments, as the resulting chunks exhibit similar statistics (e.g., line counts and token counts) to the fixed-size chunking baseline. A sensitivity analysis of \textit{max\_chunk\_size} is presented in Table~\ref{tab:cast_ablation_max_chunk_size}. We observe that retrieval and generation performance peak when \textit{max\_chunk\_size} is between 2000 and 2500 characters. Additionally, generation performance also depends on \textit{max\_context\_length}, as shown in the previous analysis. When context length allows, larger chunks can provide more information, while smaller chunks help mitigate the risk of truncation.

\begin{table}[t!]
\small
\centering
\setlength{\tabcolsep}{2pt}
\resizebox{\linewidth}{!}{
  \begin{tabular}{l|l|ccccc}
    \toprule
    & \multicolumn{1}{c|}{\multirow{2}{*}{\textbf{Metric (Model)}}} & \multicolumn{5}{c}{\textbf{Maximum chunk size}}\\
    \cmidrule(lr){3-7}
    {} & & {1000} & {1500} & {2000} & {2500} & {3000}\\
\midrule
\multirow{1}{*}{R}
    & nDCG & 69.0 & 68.4 & 71.1 & 72.3 & 69.4 \\
\midrule
\multirow{1}{*}{G}
    & \metricmodel{Pass@1}{StarCoder2} & 43.4 & 45.8 & 51.7 & 50.1 & 51.2 \\
\bottomrule
  \end{tabular}
}
\vspace{-1mm}
\caption{Ablation study of maximum chunk size effects on retrieval and generation performance.}
\label{tab:cast_ablation_max_chunk_size}
\end{table}

\section{Related Work}
\label{sec:related_work}

\paragraph{Structure-aware modeling in code tasks.}
Early work showed clear benefits from feeding explicit syntax to models: {TranX} (grammar-guided decoding) and path-based encoders {code2vec}/{code2seq} leveraged AST productions or paths to outperform token-only baselines in NL-to-code and summarization \citep{yin18tranx,alon2019code2vec,alon2019code2seq}.  
Transformer-era studies refined this idea.  GraphCodeBERT~\cite{guo2021graphcodebert} and the Code Transformer~\cite{zugner2021codetransformer} inject data-flow edges or AST distances, while CODEDISEN~\citep{zhang2021codedisen} disentangles syntax from semantics for cross-language transfer.  
More recent models layer structure-aware objectives onto large LMs: TypeT5~\citep{wei2023typet5} adds static-analysis context for type inference, and AST-T5~\cite{gong2024astt5} and StructCoder~\citep{tipirneni2024structcoder} mask or generate subtrees to boost transpilation and Java-Python translation.  

Although modern LLMs can often internalize such structure from raw tokens, these results indicate that explicit syntax still provides measurable gains—especially in preprocessing steps like chunking, where respecting function or class boundaries directly controls what the model sees. In light of the importance of structure awareness in the above literature, we propose to leverage the tree structure of code snippets to improve chunking.

\paragraph{Retrieval-augmented code generation.}
Successful code RAG hinges on pairing high-quality retrievers with generation frameworks that can effectively leverage the fetched context. 
General‐purpose systems—{RAG}~\cite{lewis2020rag}, {FiD}~\cite{izacard2021fid}, and {RePlug}~\citep{shi2023replug}—demonstrate that feeding high-recall evidence to a language model markedly improves factuality.  
In the software-engineering domain, {CodeRAG-Bench}~\citep{wang2024coderagbench} confirms these gains on repository-level tasks while revealing that lexical-matching retrievers often miss relevant code, motivating code-specific retrieval models.  
State-of-the-art code retrievers such as {CodeBERT}~\cite{feng2020codebert}, {UniXcoder}~\cite{guo2022unixcoder}, and {CodeRetriever}~\cite{li2022coderetriever} learn joint code–text or code–code embeddings and consistently surpass generic dense models in code search and question answering.  
Most pipelines still inherit fixed line-based chunking from natural-language RAG. Our work shows that respecting syntactic units with AST-aware chunks further enhances these retrieval-generation loops.

Most relevantly, CodeCRAG~\citep{du2025codegragbridginggapnatural} utilizes the graphical view of code flow to improve the overall LLM code generation pipeline. \citet{shen2024improving,xia2024improving,song-etal-2024-revisiting} propose to compute code similarity based on the graph structure of code.
In our work, we conduct a fine-grained study on one important block of code RAG workflow: chunking.
\section{Conclusion and Discussion}
\label{sec:conclusion}

In this work, we present \ourwork as a simple and effective chunking strategy for retrieval-augmented code generation. Through the structural awareness brought by AST, we are allowed to maintain syntactic integrity and high information density during chunking. Extensive experiments on various retrievers, LLM generators, and code generation tasks, validate the gain from \ourwork over the commonly used fixed-size chunking strategy on both retrieval and RAG tasks. 

By maintaining the original RAG pipeline, for the code agent practitioner, \ourwork could be used as a simple plug-and-play tool to provide informative and formatted chunks for later stage agent use. For code RAG benchmark developers, \ourwork could serve as additional resources and an effective alternative or complementary retrieval unit.


\section*{Limitations}

\paragraph{Contextual Awareness.} In our experiments, for a fair comparison, we maintain the original retrieval-augmented code generation pipeline to parse code snippets into self-contained chunks, without explicit contextual awareness from higher chunking units in the AST. However, as shown in~\cite{sarthi2024raptor,cai2024textttmixgrenhancingretrievergeneralization}, in textual RAG, including multi-level information in the tree structures can improve the retrieval performance, which can also potentially benefit code retrieval with the natural structures that can be extracted with our AST framework.

\paragraph{Multi-view of the code.} In this work, we mainly explore chunking with pure code files. However, each code snippet can potentially have multiple views, e.g., the input-output elicitation in the comments, natural language descriptions, pseudo code, and etc. Each of these views can emphasize different facets of the very code snippet. Previous work shows that including multiple views helps model math reasoning~\cite{liang2023mintboostinggeneralizationmathematical}. Similarly, instead of pure AST-based chunking on code snippets, including different chunk candidates from different views can potentially relieve the code completeness reliance of our \texttt{cAST}.

\paragraph{Inner Execution Dynamics.} In this work, we focus on introducing the structural awareness to retrieval augmented generation with AST, as a static analysis of the code semantics. However, the execution trace~\cite{ni2024nextteachinglargelanguage}, type inference~\cite{wei2023typet5}, and compilation~\cite{cummins2024metalargelanguagemodel} can potentially lead to a deep understanding of the variable dynamics. Introducing the awareness of such in-depth query analysis can help augment our \texttt{cAST} with per-query adaptiveness.

\section*{Acknowledgments}
The authors thank Jamie Callan, Fernando Diaz, Graham Neubig, Daniel Fried, and Pengcheng Yin for their insights into design and evaluation choices. The authors also thank the constructive discussions with colleagues from CMU WInE Lab and Augment Code. Xinran Zhao is supported by the ONR Award N000142312840. This work is supported by the OpenAI Research Credit program, the Amazon AI Research Gift Fund, and the Gemma Academic Program GCP Credit Award.

\bibliography{metric}

\clearpage

\appendix

\section{Appendix}
\label{sec:appendix}

\subsection{Implementation Details}
\label{sec:experiment_setup}
For Gemini and Claude models, we use the official API service. For other open-sourced models,  we use locally served models on nodes with 8 Nvidia A100 (40G) GPU and 8 Nvidia A6000 (40G) GPUs with CUDA 12 installed. Our inference structure is built upon vLLM~\citep{kwon2023efficient}. 

For fair comparison of chunks with varying sizes, instead of using top-k chunks directly, 
We use \texttt{max\_context\_length} to sequentially include retrieved chunks up to a threshold, truncating the final chunk if needed. We set the limit to 4000 for RepoEval and SWE-Bench, and extend it to 10000 for CrossCodeEval to test cross-file retrieval. 
\footnote{We use default tokenizers for open-weighted models, and \texttt{cl100k\_base} for API models.}
For generation, we adopt different settings based on evaluation metrics based on prior work~\citep{wang2024coderagbench, li2023starcoder, xia2024-agentless}: We use $t$ = 0.2, $top_p$ = 0.95, and 1 sample for Pass@1; $t$ = 0.8 and 8 samples for Pass@8.

\subsection{Metric Score Mapping Details}
\label{sec:line_mapping_metric}
In Section~\ref{subsec:exp_setup}, we denote the distributional incomparability across corpses. We implement a score mapping technique to align AST-based retrieval scores over baselines. 

Specifically, similar to~\cite{chen2023densex}, we assign each line of code a score inherited from its corresponding AST chunk. These line-level scores are then aggregated to recompute the scores of baseline chunks, allowing us to rerank them and estimate AST-based retrieval performance within the baseline framework.

\subsection{AST-based Chunking Algorithm Details}

In the main paper, we provide textual descriptions of our algorithm. Here, we present the pseudo code of our implementation in Alg. \ref{alg:cast}.

\begin{algorithm}[t]
\small
\caption{AST-based Chunking Algorithm}
\label{alg:cast}
\begin{algorithmic}[1]
\State MAX\_SIZE $\gets$ maximum chunk size
\\
\Function{ChunkCode}{code}
    \State tree $\gets$ \Call{ParseAST}{code}
    \If{\Call{GetSize}{code} $\leq$ MAX\_SIZE}
        \State \Return $[\text{tree}]$
    \Else
        \State \Return \Call{ChunkNodes}{tree.children}
    \EndIf
\EndFunction
\\
\Function{ChunkNodes}{nodes}
    \State chunks $\gets$ [ ], chunk $\gets$ [ ], size $\gets$ 0
    \For{node in nodes}
        \State s $\gets$ \Call{GetSize}{node}
        \If{(chunk = [ ] and s $>$ MAX\_SIZE) or\\
            \hspace{2.6em} (size + s $>$ MAX\_SIZE)} 
            \If{chunk $\neq$ [ ]}
                \State chunks.append(chunk)
                \State chunk, size $\gets$ [ ], 0
            \EndIf
            \If{s $>$ MAX\_SIZE}
                \State subchunks $\gets$ \Call{ChunkNodes}{node.children}
                \State chunks.extend(subchunks)
                \State \textbf{continue}
            \EndIf
        \Else
            \State chunk.append(node); size $\gets$ size + s
        \EndIf
    \EndFor
    \If{chunk $\neq$ [ ]}
        \State chunks.append(chunk)
    \EndIf
    \State \Return chunks
\EndFunction
\end{algorithmic}
\end{algorithm}

\begin{table*}[t]
\vspace{-4mm}
\centering
\resizebox{\textwidth}{!}{
  \begin{tabular}{l|cccccc|cccccc}
    \toprule
    \multicolumn{1}{c|}{\multirow{2}{*}{\textbf{Method}}} & \multicolumn{6}{c|}{\textbf{\ourwork}} & \multicolumn{6}{c}{\textbf{Fixed-size}} \\
    {} & {nDCG@5} & {nDCG@10} & {P@5} & {P@10} & {Recall@5} & {Recall@10} & {nDCG@5} & {nDCG@10} & {P@5} & {P@10} & {Recall@5} & {Recall@10}\\
    \midrule
    \multicolumn{13}{c}{\textit{RepoEval}} \\
    \midrule
BGE-base          
& 71.1 & 74.7 & \good{34.9} & \good{20.4} & \good{69.8} & \better{77.6} & 71.3 & 74.6 & 32.8 & 19.1 & 67.4 & 74.1
\\
BGE-large         
& \good{72.2} & \good{75.4} & \better{34.9} & \good{20.2} & \better{69.6} & \better{76.3} & 71.1 & 73.9 & 31.3 & 18.1 & 64.9 & 70.6
\\
GIST-base         
& \good{75.9} & 78.5 & \better{38.1} & 21.2 & \better{75.0} & \good{80.5} & 74.2 & 78.0 & 34.8 & 20.6 & 70.7 & 78.5
\\
GIST-large        
& \better{78.9} & \good{81.9} & \better{38.8} & 22.0 & \better{76.6} & \good{82.8} & 75.1 & 79.5 & 34.8 & 21.1 & 71.1 & 80.2
\\
Codesage-small-v2
& \good{85.1} & \good{88.8} & \good{44.1} & 25.3 & \good{83.9} & \good{91.0} & 83.0 & 86.4 & 42.9 & 24.5 & 82.1 & 89.1
\\
Jina-v2-code      
& 87.1 & 90.5 & \good{47.9} & 27.1 & \good{87.9} & \good{94.7} & 86.8 & 90.9 & 46.3 & 26.7 & 84.9 & 92.9
\\
\midrule
\multicolumn{13}{c}{\textit{SWE-bench}} \\
\midrule
BGE-base & \good{44.0} & \good{41.5} & \good{39.7} & \good{32.5} & \good{18.4} & \good{26.8} & 42.4 & 39.5 & 38.3 & 31.2 & 17.3 & 24.4 \\
BGE-large & 42.2 & 40.4 & 37.7 & 31.6 & 17.5 & \good{26.1} & 42.8 & 39.9 & 38.3 & 31.2 & 17.0 & 24.6 \\
GIST-base & \good{44.4} & \good{42.5} & 39.1 & \good{32.9} & 18.5 & \good{27.6} & 43.1 & 40.6 & 38.6 & 31.8 & 17.8 & 25.9 \\
GIST-large & 44.0 & 41.9 & 39.5 & 33.1 & 18.5 & 27.0 & 43.5 & 41.7 & 39.2 & 33.2 & 18.0 & 26.5 \\
Codesage-small-v2 & 43.1 & \good{41.4} & \good{38.8} & \good{32.8} & 18.3 & \good{26.4} & 42.6 & 40.0 & 37.5 & 31.0 & 17.5 & 24.7 \\
    \bottomrule
  \end{tabular}
}
\vspace{-1mm}
\caption{Retrieval performance (nDCG, Precision, Recall@\{5,10\}) on RepoEval and SWE-bench.}
\label{tab:full_retrieval_results}
\vspace{3mm}
\end{table*}

\begin{table*}[t]
\small
\centering
  \begin{tabular}{l|cc|cc}
    \toprule
    \multicolumn{1}{c|}{\multirow{2}{*}{\textbf{Method}}} & \multicolumn{2}{c|}{\textbf{\ourwork}} & \multicolumn{2}{c}{\textbf{Fixed-size}} \\
    {} & {StarCoder2} & {CodeLlama} & {StarCoder2} & {CodeLlama} \\
    \midrule
BGE-base   & \better{51.7} & \better{49.6} & 47.5 & 45.6 \\
BGE-large  & \better{48.8} & \good{50.9} & 45.8 & 49.9 \\
GIST-base  & \better{57.9} & \better{56.6} & 51.2 & 51.5 \\
GIST-large & \good{61.7} & \better{60.3} & 59.2 & 55.5 \\
Codesage-small-v2 & \better{73.2} & \better{72.1} & 67.6 & 66.5 \\
Jina-v2-code & \better{80.7} & \good{75.9} & 75.1 & 75.1 \\
    \bottomrule
  \end{tabular}
\caption{RAG performance (Pass@1) on RepoEval with various retrievers.}
\label{tab:full_repoeval_results}
\vspace{-1mm}
\end{table*}

\begin{table*}[t]
\vspace{-1mm}
\small
\centering
  \begin{tabular}{l|cc|cc}
    \toprule
    \multicolumn{1}{c|}{\multirow{2}{*}{\textbf{Method}}} & \multicolumn{2}{c|}{\textbf{\ourwork}} & \multicolumn{2}{c}{\textbf{Fixed-size}} \\
    {} & {Claude-3.7-Sonnet} & {Gemini-2.5-pro} & {Claude-3.7-Sonnet} & {Gemini-2.5-pro}\\
    \midrule
BGE-base   & \better{16.3} & \better{35.3} &13.7 & 32.3  \\
BGE-large  & 13.3 & 30.3 & 14.6 & 33.7  \\
GIST-base  & 15.0 & 33.7 & 14.7 & 33.0  \\
GIST-large   & \better{15.3} & 31.0 & 13.0  & 33.0 \\
Codesage-small-v2 & \better{16.7} & \better{32.7} & 14.0 & 31.0  \\
    \bottomrule
  \end{tabular}
\vspace{-1mm}
\caption{RAG performance (Claude w/ Pass@1 \& Gemini w/ Pass@8) on SWE-bench.}
\label{tab:full_swebench_results}
\vspace{-1mm}
\end{table*}

\begin{table*}[t]
\footnotesize  
\centering
\setlength{\tabcolsep}{4pt} 
  \begin{tabular}{l|cccc|cccc}
    \toprule
    \multicolumn{1}{c|}{\multirow{2}{*}{\textbf{Method}}} & \multicolumn{4}{c|}{\textbf{\ourwork}} & \multicolumn{4}{c}{\textbf{Fixed-size}} \\
    {} & {EM (code)} & {ES (code)} & {EM (id)} & {F1 (id)} & {EM (code)} & {ES (code)} & {EM (id)} & {F1 (id)}\\
    \midrule
    \multicolumn{9}{c}{\textit{BGE-base + Starcoder2-7B}} \\
    \midrule
Python      & \good{23.8} & \good{72.2} & \good{34.7} & \good{63.8} & 21.2 & 71.0 & 32.0 & 62.1 \\
Java        & 27.8 & 70.9 & 37.5 & 63.8 & 27.3 & 71.6 & 37.1 & 64.1 \\
C\#         & \good{26.9} & \good{73.5} & \better{32.0} & \good{56.4} & 23.9 & 71.8 & 28.3 & 53.8 \\
TypeScript  & \good{13.4} & \better{49.6} & \good{19.5} & \better{43.6} & 11.4 & 46.0 & 17.4 & 40.2 \\
    \midrule
    \multicolumn{9}{c}{\textit{GIST-base + Starcoder2-7B}} \\
    \midrule
Python      & 23.4 & 71.9 & 34.0 & 63.7 & 23.0 & 71.7 & 33.5 & 63.3 \\
Java        & \good{28.0} & 71.2 & 37.7 & 64.3 & 27.0 & 71.3 & 36.8 & 63.7 \\
C\#         & \good{26.6} & 73.2 & \good{31.2} & \good{56.0} & 24.3 & 72.5 & 28.7 & 54.3 \\
TypeScript  & \good{13.0} & \better{49.3} & \good{19.7} & \better{43.9} & 11.2 & 46.1 & 17.2 & 40.2 \\
    \midrule
    \multicolumn{9}{c}{\textit{Codesage-small-v2 + Starcoder2-7B}} \\
    \midrule
Python      & \better{29.1} & \good{74.3} & \better{39.9} & \good{67.6} & 24.8 & 73.1 & 36.3 & 65.7 \\
Java        & \good{30.9} & 72.2 & \good{41.2} & \good{66.1} & 28.1 & 71.5 & 38.3 & 64.6 \\
C\#         & \good{28.3} & \good{74.2} & \better{33.4} & \better{58.2} & 25.5 & 72.4 & 29.9 & 54.9 \\
TypeScript  & \good{13.7} & \better{49.1} & \good{19.6} & \good{43.5} & 11.9 & 46.0 & 17.7 & 40.6 \\
    \bottomrule
  \end{tabular}
\vspace{-1mm}
\caption{RAG performance (Code Match \& Identifier Match) on CrossCodeEval. }
\label{tab:full_cceval_results}
\vspace{-3mm}
\end{table*}

\subsection{Extended Experiment Results}
\label{sec:extended_results}

In the main paper, we show concise results from our experiment to demonstrate a clear contribution. We further include detailed results from our settings here.
In Table~\ref{tab:full_retrieval_results}, we present the retrieval performance with various metrics and retrievers on RepoEval and SWE-bench. 
In Table~\ref{tab:full_swebench_results}, we present the RAG performance on SWE-Bench with various retrievers (large language models) and generators. 
In Table~\ref{tab:full_repoeval_results}, we present the RAG performance on RepoEval with various retrievers and generators. 
In Table~\ref{tab:full_cceval_results}, we show the RAG performance with various retrievers on CCEval across different programming languages.s

These tables show similar conclusions with our findings in the main paper, where \ourwork consistently performs better than fixed-size line-based chunking with syntactic integrity and high information density.

\subsection{Performance differences across different programming languages} 

A key limitation of fixed-size, line-based chunking is its poor generalizability across programming languages. Language-specific syntax means a line limit tuned for one language over- or under-segments another, leading to uneven information density and degraded retrieval and generation quality. In contrast, \ourwork uses structure-aware segmentation based on abstract-syntax units common across languages, mitigating these issues.

Table \ref{tab:full_cceval_results} reports results with the Codesage-small-v2 + Starcoder2-7B pipeline. Though both methods use fixed chunk lengths, performance variation across languages is notably higher for the baseline. Averaged over four languages, \ourwork improves EM by 2.9 on code and 3.0 on identifier, with the largest gains on TypeScript—the noisiest language. These consistent gains highlight the value of respecting syntax when handling multilingual code.

The performance differences across different languages with different chunking strategies, as well as RAG design choices, can form an interesting future line of work.


\subsection{Ethical Statements}
We foresee no ethical concerns or potential risks in our work.
All of the retrieval models, code generators, and datasets are open-sourced or with public APIs, as shown in Section~\ref{sec:experiment}. 
The LLMs we applied in the experiments are also publicly available.
Given our context, the outputs of LLMs (code snippets) are unlikely to contain harmful and dangerous information. All the code is executed in sandboxes, with no threat to the public internet. The natural language part of our experiments is mainly on English. Multiple programming languages are included: Python, Java, C\#, and TypeScript.

Our code is open source and available at \url{https://github.com/yilinjz/astchunk}.

\subsection{Licenses of scientific artifacts}
We conclude the licenses of the scientific artifacts we used in Table~\ref{tab:tools}. All of our usage for scientific discovery follows the original purpose of the artifacts.

\begin{table*} [t]
    \centering
    \resizebox{\textwidth}{!}{
    \begin{tabular}{llll} 
       \toprule
        \textbf{Artifacts/Packages} & \textbf{Citation} & \textbf{Link} & \textbf{License}\\ 
        \midrule
        RepoEval & \citep{zhang-2023-repoeval} & \url{https://github.com/irgroup/repro_eval} & MIT License \\
        SWE-bench & \citep{jimenez-2024-swebench}  & \url{https://github.com/SWE-bench/SWE-bench} & MIT License \\
        CrossCodeEval \ & \cite{ding-2023-crosscodeeval} & \url{https://github.com/amazon-science/cceval} & Apache License 2.0 \\

        \midrule
        PyTorch & \cite{paszke-etal-2019-pytorch} & \url{https://pytorch.org/} & BSD-3 License\\
        transformers & \cite{wolf2019huggingface} & \url{https://huggingface.co/transformers/v2.11.0/index.html} & Apache License 2.0\\
        numpy & \cite{DBLP:journals/nature/HarrisMWGVCWTBS20} & \url{https://numpy.org/} & BSD License \\
        matplotlib & \cite{hunter2007matplotlib} & \url{https://matplotlib.org/} & BSD compatible License\\
        vllm & \cite{kwon2023efficient} & \url{https://github.com/vllm-project/vllm} & Apache License 2.0 \\
       \midrule
        BGE & \citep{bge_embedding}  & \url{https://huggingface.co/BAAI/bge-large-en} & MIT license\\
        GIST & \citep{solatorio2024gistembed} & \url{https://huggingface.co/avsolatorio/GIST-Embedding-v0} & MIT license \\
        CodeSage & \citep{zhang2024codesage} & \url{https://huggingface.co/codesage/codesage-small-v2} & Apache License 2.0 \\
        Jina-v2-Code & \citep{gunther2023jina} & \url{https://huggingface.co/jinaai/jina-embeddings-v2-base-code} & Apache License 2.0 \\
        StarCoder2 & \citep{lozhkov2024starcoder} & \url{https://huggingface.co/bigcode/starcoder2-7b} & \href{https://www.bigcode-project.org/docs/pages/bigcode-openrail/}{LICENSE} \\ 
        CodeLlama & \citep{roziere2023code}& \url{https://huggingface.co/codellama/CodeLlama-7b-hf} &  \href{https://huggingface.co/meta-llama/Llama-2-7b-chat-hf/blob/main/LICENSE.txt}{LICENSE}\\
      \bottomrule
    \end{tabular}}
    \caption{Details of datasets, major packages, and existing models we use. The curated datasets and our code/software are under the MIT License.}
    \label{tab:tools}
\end{table*}

\end{document}